\documentclass[sigconf]{acmart}
\usepackage{balance}

\AtBeginDocument{%
  \providecommand\BibTeX{{%
    \normalfont B\kern-0.5em{\scshape i\kern-0.25em b}\kern-0.8em\TeX}}}

\copyrightyear{2024}
\acmYear{2024}
\setcopyright{rightsretained}
\acmConference[MM '24]{Proceedings of the 32nd ACM International Conference
on Multimedia}{October 28-November 1, 2024}{Melbourne, VIC, Australia}
\acmBooktitle{Proceedings of the 32nd ACM International Conference on
Multimedia (MM '24), October 28-November 1, 2024, Melbourne, VIC, Australia}
\acmDOI{10.1145/3664647.3685000}
\acmISBN{979-8-4007-0686-8/24/10}

\begin{document}

\title{Muskits-ESPnet: A Comprehensive Toolkit for \\Singing Voice Synthesis in New Paradigm}



\author{Yuning Wu}
\affiliation{%
  \institution{Renmin University of China}
 \city{Beijing}
 \country{China}
  }
\email{yuningwu@ruc.edu.cn}

\author{Jiatong Shi}
\affiliation{%
  \institution{Carnegie Mellon University}
 \city{Pittsburgh}
 \country{United States}
  }
\email{jiatongs@cs.cmu.edu}

\author{Yifeng Yu}
\affiliation{%
  \institution{Georgia Institute of Technology}
 \city{Atlanta}
 \country{United States}
  }
\email{yyu479@gatech.edu}

\author{Yuxun Tang}
\affiliation{%
  \institution{Renmin University of China}
 \city{Beijing}
 \country{China}
  }
\email{tangyuxun@ruc.edu.cn}

\author{Tao Qian}
\affiliation{%
  \institution{Renmin University of China}
 \city{Beijing}
 \country{China}
  }
\email{qiantao@shsid.org}

\author{Yueqian Lin}
\affiliation{%
  \institution{Duke Kunshan University}
 \city{Durham}
 \country{United States}
  }
\email{yueqian.lin@duke.edu}

\author{Jionghao Han}
\affiliation{%
  \institution{Carnegie Mellon University}
 \city{Pittsburgh}
 \country{United States}
  }
\email{jionghah@andrew.cmu.edu}

\author{Xinyi Bai}
\affiliation{%
  \institution{Cornell University}
 \city{Ithaca}
 \country{United States}
  }
\email{xb@cornell.edu}


\author{Shinji Watanabe}
\affiliation{%
  \institution{Carnegie Mellon University}
 \city{Pittsburgh}
 \country{United States}
  }
\email{shinjiw@ieee.org}

\author{Qin Jin}
\affiliation{%
  \institution{Renmin University of China}
 \city{Beijing}
 \country{China}
  }
\email{qjin@ruc.edu.cn}

\renewcommand{\shortauthors}{Yuning Wu et al.}



\begin{abstract}
This research presents Muskits-ESPnet, a versatile toolkit that introduces new paradigms to Singing Voice Synthesis (SVS) through the application of pretrained audio models in both continuous and discrete approaches. Specifically, we explore discrete representations derived from SSL models and audio codecs and offer significant advantages in versatility and intelligence, supporting multi-format inputs and adaptable data processing workflows for various SVS models. The toolkit features automatic music score error detection and correction, as well as a perception auto-evaluation module to imitate human subjective evaluating scores.
Muskits-ESPnet is 
available at \url{https://github.com/espnet/espnet}.
\end{abstract}





\maketitle

\section{Introduction}
\label{sec:intro}

\begin{figure}[t]
	\centering
	\includegraphics[width=1\columnwidth]{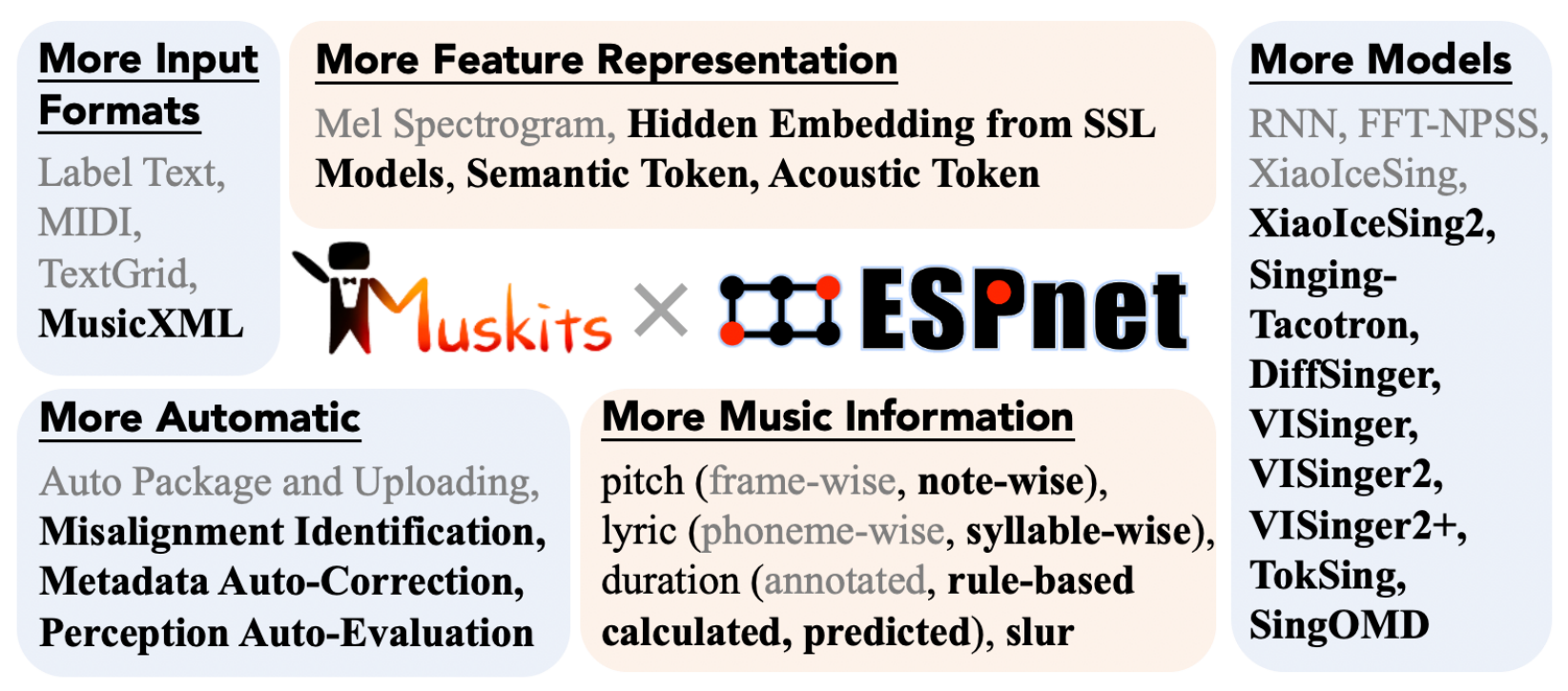}
        \vspace{-20pt}
	\caption{\small Improvements of Muskits-ESPnet compared with its origin version. The boldface indicates new functions.}
 \vspace{-15pt}
	\label{fig:version}
\end{figure}

SVS converts music scores into vocal singing using a specific singer's voice, aiming for accurate lyrics, pitch, and duration while ensuring a realistic sound. It faces challenges in achieving high standards of pitch, prosody, and emotional expression due to complex data processing requirements.

\begin{figure*}[t]
	\centering
	\includegraphics[width=1\textwidth]{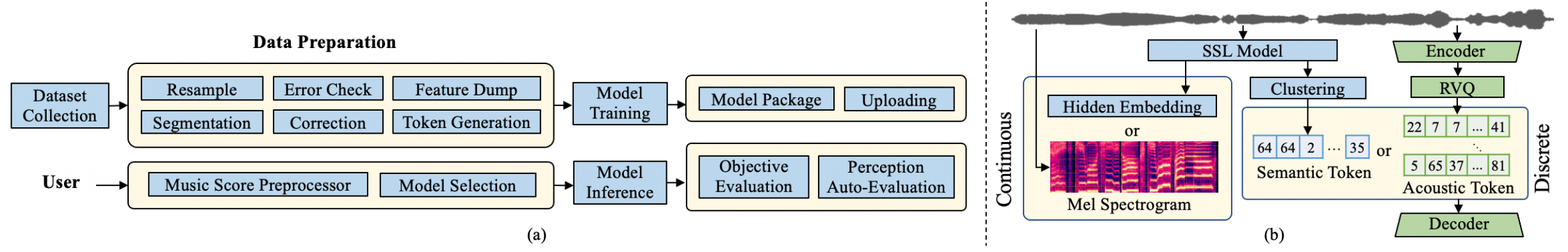}
        \vspace{-15pt}
	\caption{\small (a) SVS workflow. The upper section illustrates the training pipeline, while the lower section shows the inference pipeline for users. The functions in the yellow blocks can be flexibly selected based on specific requirements.(b) Different SVS feature representations. Continuous features include mel spectrograms and hidden embeddings from SSL models. Discrete representations consist of semantic tokens clustered from SSL models and acoustic tokens extracted from codecs.}
	\label{fig:flow}
 \vspace{-13pt}
\end{figure*}

The common approach for SVS~\cite{Shi2020SequenceToSequenceSV, Lu2020XiaoiceSingAH, gu2021bytesing, wang2022xiaoicesing, guo2022singaug} involves an acoustic model predicting acoustic feature representations from music scores, followed by a vocoder~\cite{Kumar2019MelGANGA, Mustafa2020StyleMelGANAE, kong2020hifi, Lee2022BigVGANAU} reconstructing audio from these features. Most music processing toolkits~\cite{sinsy, nnsvs} for SVS, including our initial version of Muskits~\cite{shi2022muskits}, follow this framework. 
However, the emergence of audio pre-training and the shift towards discrete representations in large models have brought new possibilities for SVS.
Previously, 80-dimensional real-valued mel-spectrograms are commonly used as acoustic representations. Now, outputs from audio pretrained models~\cite{Baevski2019vqwav2vecSL, baevski2020wav2vec, Hsu2021HuBERTSS, Chen2021WavLMLS, Li2023MERTAM, shi2023multi} trained on large-scale datasets can assist acoustic modeling or extract discrete representations~\cite{hayashi2020discretalk,polyak21_interspeech, lee2022direct, shi2023multi, shi2023enhancing, shi2023bridging, chang23b_interspeech, chang2023exploring, barrault2023seamless, yan-etal-2023-espnet, yang2023towards}. This approach efficiently meets the need for data discretization with large models~\cite{zhang2023speechgpt, wang2023neural, anastassiou2024seed}. Our work focuses on these new SVS paradigms and optimizes the entire data flow accordingly.

Our Muskits-ESPnet toolkit demonstrates exceptional versatility and intelligence (see Figure~\ref{fig:version}). We enhance the SVS models by integrating pre-trained models with traditional continuous feature-based approaches and introducing a new paradigm based on discrete representations. Furthermore, the entire data processing workflow is optimized to support all music file formats, not just specific datasets, and includes an automatic error-check and correction module to improve data alignment accuracy. We compile common feature representations to accommodate different SVS model inputs and introduce a perception auto-evaluation model~\cite{singmos}, significantly reducing the cost and effort of manual scoring. Our Muskits-ESPnet toolkit supports the most advanced SVS models and automates the entire data processing workflow (see Figure~\ref{fig:flow}). Recently, our toolkit serves as the baseline for the SVS track in Interspeech 2024 Discrete Speech Unit Challenge~\cite{chang2024interspeech}.

\section{New Paradigms in SVS}
\label{sec:sys}

Advances in audio pretraining technology impact audio generation tasks significantly. We apply this to SVS in two ways:

First, we enhance traditional SVS models by integrating pretrained audio encodings, replacing or complementing mel spectrograms (see Figure~\ref{fig:flow} (b)). Our new SVS model~\cite{yu2024visinger2}, based on a Variational Auto-Encoder~\cite{Zhang2022VISinger2H}, performs better with joint encoding than with spectrograms alone.

Second, we explore SVS using discrete representations from pretrained models, including semantic tokens from SSL model outputs and acoustic tokens from an audio codec~\cite{Zeghidour2021SoundStreamAE, Defossez2022HighFN}. Our discrete-based SVS models~\cite{wu2024toksing, tang2024singomd} in ESPnet achieve lower spatial costs compared to continuous representations.

\section{Implementations}
\label{sec:imp}

The Muskits-ESPnet data flow, illustrated in Figure~\ref{fig:flow} (a), includes resampling, segmenting, error correction, and feature computation during training. Post-training, the model is packaged for upload. For inference, user inputs are preprocessed, the SVS model is selected, and evaluations are performed. Our framework supports various data types and model configurations, offering flexible functionality based on user needs. Detailed procedures for data preparation, training, inference, and evaluation are provided.

\subsection{Data Preparation}
This stage involves preprocessing raw music data into input sequences for SVS models. Typically, we extract sequences of three essential elements: <lyrics, pitch, duration> from various formats such as MusicXML, MIDI, and TextGrid. Upon reviewing several datasets, we identify a notable percentage of annotation errors, including redundant, missing, or misalignment of lyrics and notes. To tackle these issues, we develop a misalignment detection module and a metadata auto-correction module with specific adaptations for different languages. Our toolkit ensures annotation alignment consistency, thereby significantly enhancing model performance~\cite{wu2023phoneix}.

\vspace{-5pt}
\subsection{Training and Inference}
Model training and inference follow the ESPnet~\cite{Watanabe2018ESPnetES} task processing workflow, supporting multi-GPU training and dynamic batching. We have significantly enhanced the generalizability of learning methods for SVS. This includes enriching joint training and fine-tuning paradigms for acoustic models and vocoders~\cite{Wu2023ASE}, and supporting both autoregressive~\cite{Wang2022SingingTacotronGD} and non-autoregressive~\cite{wang2022xiaoicesing, Lu2020XiaoiceSingAH, Shi2020SequenceToSequenceSV, Blaauw2019SequencetoSequenceSS, Liu2021DiffSingerSV, Zhang2021VISingerVI, Zhang2022VISinger2H, yu2024visinger2, wu2024toksing, tang2024singomd} acoustic prediction methods. Additionally, the vocoder section now accommodates both continuous and discrete representations and includes transfer learning workflows~\cite{wu2024toksing, tang2024singomd}. We have also optimized the data processing workflow, ensuring compatibility with different models while reducing time costs by approximately 60\% compared to the previous generation.

\vspace{-5pt}
\subsection{Evaluation}
We employ a comprehensive set of objective metrics to evaluate the similarity between generated audio and the original audio across various dimensions, including Mel Cepstral Distortion (MCD), Root Mean Square Error of Fundamental Frequency (F0\_RMSE), Semitone Accuracy (SA), and Voiced/Unvoiced Error Rate (VUV\_E). For listening feelings, we introduce an innovative perception auto-evaluation module~\cite{singmos} to emulate human judgment.

\section{Conclusion}
\label{sec:conclusion}


Muskits-ESPnet advances SVS by integrating audio pretraining and exploring both continuous and discrete representations, enhancing model capability and efficiency. It features robust data preprocessing, error correction, and support for diverse inputs. Optimized training and inference workflows, along with auto-evaluation, demonstrate its potential to support cutting-edge SVS models while reducing costs, setting a new standard for future SVS developments.

\begin{acks}
This work was partially supported by the  National Natural Science Foundation of China (No. 62072462).
\end{acks}
\bibliographystyle{ACM-Reference-Format}
\balance
\bibliography{mybib}


\begin{thebibliography}{46}


\ifx \showCODEN    \undefined \def \showCODEN     #1{\unskip}     \fi
\ifx \showDOI      \undefined \def \showDOI       #1{#1}\fi
\ifx \showISBNx    \undefined \def \showISBNx     #1{\unskip}     \fi
\ifx \showISBNxiii \undefined \def \showISBNxiii  #1{\unskip}     \fi
\ifx \showISSN     \undefined \def \showISSN      #1{\unskip}     \fi
\ifx \showLCCN     \undefined \def \showLCCN      #1{\unskip}     \fi
\ifx \shownote     \undefined \def \shownote      #1{#1}          \fi
\ifx \showarticletitle \undefined \def \showarticletitle #1{#1}   \fi
\ifx \showURL      \undefined \def \showURL       {\relax}        \fi
\providecommand\bibfield[2]{#2}
\providecommand\bibinfo[2]{#2}
\providecommand\natexlab[1]{#1}
\providecommand\showeprint[2][]{arXiv:#2}

\bibitem[Anastassiou et~al\mbox{.}(2024)]%
        {anastassiou2024seed}
\bibfield{author}{\bibinfo{person}{Philip Anastassiou}, \bibinfo{person}{Jiawei Chen}, \bibinfo{person}{Jitong Chen}, \bibinfo{person}{Yuanzhe Chen}, \bibinfo{person}{Zhuo Chen}, \bibinfo{person}{Ziyi Chen}, \bibinfo{person}{Jian Cong}, \bibinfo{person}{Lelai Deng}, \bibinfo{person}{Chuang Ding}, \bibinfo{person}{Lu Gao}, {et~al\mbox{.}}} \bibinfo{year}{2024}\natexlab{}.
\newblock \showarticletitle{Seed-TTS: A Family of High-Quality Versatile Speech Generation Models}.
\newblock \bibinfo{journal}{\emph{arXiv preprint arXiv:2406.02430}} (\bibinfo{year}{2024}).
\newblock


\bibitem[Baevski et~al\mbox{.}(2019)]%
        {Baevski2019vqwav2vecSL}
\bibfield{author}{\bibinfo{person}{Alexei Baevski}, \bibinfo{person}{Steffen Schneider}, {and} \bibinfo{person}{Michael Auli}.} \bibinfo{year}{2019}\natexlab{}.
\newblock \showarticletitle{vq-wav2vec: Self-Supervised Learning of Discrete Speech Representations}. In \bibinfo{booktitle}{\emph{ICLR}}.
\newblock


\bibitem[Baevski et~al\mbox{.}(2020)]%
        {baevski2020wav2vec}
\bibfield{author}{\bibinfo{person}{Alexei Baevski}, \bibinfo{person}{Yuhao Zhou}, \bibinfo{person}{Abdelrahman Mohamed}, {et~al\mbox{.}}} \bibinfo{year}{2020}\natexlab{}.
\newblock \showarticletitle{wav2vec 2.0: A framework for self-supervised learning of speech representations}.
\newblock \bibinfo{journal}{\emph{NeurIPS}} (\bibinfo{year}{2020}).
\newblock


\bibitem[Barrault et~al\mbox{.}(2023)]%
        {barrault2023seamless}
\bibfield{author}{\bibinfo{person}{Lo{\"\i}c Barrault}, \bibinfo{person}{Yu-An Chung}, \bibinfo{person}{Mariano~Coria Meglioli}, \bibinfo{person}{David Dale}, \bibinfo{person}{Ning Dong}, \bibinfo{person}{Mark Duppenthaler}, \bibinfo{person}{Paul-Ambroise Duquenne}, \bibinfo{person}{Brian Ellis}, \bibinfo{person}{Hady Elsahar}, \bibinfo{person}{Justin Haaheim}, {et~al\mbox{.}}} \bibinfo{year}{2023}\natexlab{}.
\newblock \showarticletitle{Seamless: Multilingual Expressive and Streaming Speech Translation}.
\newblock \bibinfo{journal}{\emph{arXiv preprint arXiv:2312.05187}} (\bibinfo{year}{2023}).
\newblock


\bibitem[Blaauw and Bonada(2019)]%
        {Blaauw2019SequencetoSequenceSS}
\bibfield{author}{\bibinfo{person}{Merlijn Blaauw} {and} \bibinfo{person}{Jordi Bonada}.} \bibinfo{year}{2019}\natexlab{}.
\newblock \showarticletitle{Sequence-to-Sequence Singing Synthesis Using the Feed-Forward Transformer}.
\newblock \bibinfo{journal}{\emph{ICASSP}} (\bibinfo{year}{2019}), \bibinfo{pages}{7229--7233}.
\newblock


\bibitem[Chang et~al\mbox{.}(2024)]%
        {chang2024interspeech}
\bibfield{author}{\bibinfo{person}{Xuankai Chang}, \bibinfo{person}{Jiatong Shi}, \bibinfo{person}{Jinchuan Tian}, \bibinfo{person}{Yuning Wu}, \bibinfo{person}{Yuxun Tang}, \bibinfo{person}{Yihan Wu}, \bibinfo{person}{Shinji Watanabe}, \bibinfo{person}{Yossi Adi}, \bibinfo{person}{Xie Chen}, {and} \bibinfo{person}{Qin Jin}.} \bibinfo{year}{2024}\natexlab{}.
\newblock \showarticletitle{The Interspeech 2024 Challenge on Speech Processing Using Discrete Units}. In \bibinfo{booktitle}{\emph{Interspeech}}.
\newblock


\bibitem[Chang et~al\mbox{.}(2023a)]%
        {chang2023exploring}
\bibfield{author}{\bibinfo{person}{Xuankai Chang}, \bibinfo{person}{Brian Yan}, \bibinfo{person}{Kwanghee Choi}, \bibinfo{person}{Jeeweon Jung}, \bibinfo{person}{Yichen Lu}, \bibinfo{person}{Soumi Maiti}, \bibinfo{person}{Roshan Sharma}, \bibinfo{person}{Jiatong Shi}, \bibinfo{person}{Jinchuan Tian}, \bibinfo{person}{Shinji Watanabe}, {et~al\mbox{.}}} \bibinfo{year}{2023}\natexlab{a}.
\newblock \showarticletitle{Exploring speech recognition, translation, and understanding with discrete speech units: A comparative study}. In \bibinfo{booktitle}{\emph{ICASSP}}.
\newblock


\bibitem[Chang et~al\mbox{.}(2023b)]%
        {chang23b_interspeech}
\bibfield{author}{\bibinfo{person}{Xuankai Chang}, \bibinfo{person}{Brian Yan}, \bibinfo{person}{Yuya Fujita}, \bibinfo{person}{Takashi Maekaku}, {and} \bibinfo{person}{Shinji Watanabe}.} \bibinfo{year}{2023}\natexlab{b}.
\newblock \showarticletitle{{Exploration of Efficient End-to-End ASR using Discretized Input from Self-Supervised Learning}}. In \bibinfo{booktitle}{\emph{Interspeech}}.
\newblock


\bibitem[Chen et~al\mbox{.}(2021)]%
        {Chen2021WavLMLS}
\bibfield{author}{\bibinfo{person}{Sanyuan Chen}, \bibinfo{person}{Chengyi Wang}, \bibinfo{person}{Zhengyang Chen}, {et~al\mbox{.}}} \bibinfo{year}{2021}\natexlab{}.
\newblock \showarticletitle{Wav{LM}: Large-Scale Self-Supervised Pre-Training for Full Stack Speech Processing}.
\newblock \bibinfo{journal}{\emph{IJSTSP}}  \bibinfo{volume}{16} (\bibinfo{year}{2021}), \bibinfo{pages}{1505--1518}.
\newblock


\bibitem[D'efossez et~al\mbox{.}(2022)]%
        {Defossez2022HighFN}
\bibfield{author}{\bibinfo{person}{Alexandre D'efossez}, \bibinfo{person}{Jade Copet}, \bibinfo{person}{Gabriel Synnaeve}, {et~al\mbox{.}}} \bibinfo{year}{2022}\natexlab{}.
\newblock \showarticletitle{High Fidelity Neural Audio Compression}.
\newblock \bibinfo{journal}{\emph{ArXiv}}  \bibinfo{volume}{abs/2210.13438} (\bibinfo{year}{2022}).
\newblock


\bibitem[gil Lee et~al\mbox{.}(2022)]%
        {Lee2022BigVGANAU}
\bibfield{author}{\bibinfo{person}{Sang gil Lee}, \bibinfo{person}{Wei Ping}, \bibinfo{person}{Boris Ginsburg}, {et~al\mbox{.}}} \bibinfo{year}{2022}\natexlab{}.
\newblock \showarticletitle{Big{VGAN}: A Universal Neural Vocoder with Large-Scale Training}.
\newblock \bibinfo{journal}{\emph{ArXiv}}  \bibinfo{volume}{abs/2206.04658} (\bibinfo{year}{2022}).
\newblock


\bibitem[Gu et~al\mbox{.}(2021)]%
        {gu2021bytesing}
\bibfield{author}{\bibinfo{person}{Yu Gu}, \bibinfo{person}{Xiang Yin}, \bibinfo{person}{Yonghui Rao}, {et~al\mbox{.}}} \bibinfo{year}{2021}\natexlab{}.
\newblock \showarticletitle{Bytesing: A {Chinese} singing voice synthesis system using duration allocated encoder-decoder acoustic models and {WaveRNN} vocoders}. In \bibinfo{booktitle}{\emph{ISCSLP}}.
\newblock


\bibitem[Guo et~al\mbox{.}(2022)]%
        {guo2022singaug}
\bibfield{author}{\bibinfo{person}{Shuai Guo}, \bibinfo{person}{Jiatong Shi}, \bibinfo{person}{Tao Qian}, {et~al\mbox{.}}} \bibinfo{year}{2022}\natexlab{}.
\newblock \showarticletitle{Sing{A}ug: Data Augmentation for Singing Voice Synthesis with Cycle-consistent Training Strategy}. In \bibinfo{booktitle}{\emph{Interspeech}}.
\newblock


\bibitem[Hayashi et~al\mbox{.}(2020)]%
        {hayashi2020discretalk}
\bibfield{author}{\bibinfo{person}{Tomoki Hayashi} {et~al\mbox{.}}} \bibinfo{year}{2020}\natexlab{}.
\newblock \showarticletitle{Discretalk: Text-to-speech as a machine translation problem}.
\newblock \bibinfo{journal}{\emph{arXiv preprint arXiv:2005.05525}} (\bibinfo{year}{2020}).
\newblock


\bibitem[Hsu et~al\mbox{.}(2021)]%
        {Hsu2021HuBERTSS}
\bibfield{author}{\bibinfo{person}{Wei-Ning Hsu}, \bibinfo{person}{Benjamin Bolte}, \bibinfo{person}{Yao-Hung~Hubert Tsai}, {et~al\mbox{.}}} \bibinfo{year}{2021}\natexlab{}.
\newblock \showarticletitle{{HuBERT}: Self-Supervised Speech Representation Learning by Masked Prediction of Hidden Units}.
\newblock \bibinfo{journal}{\emph{TASLP}}  \bibinfo{volume}{29} (\bibinfo{year}{2021}), \bibinfo{pages}{3451--3460}.
\newblock


\bibitem[Kong et~al\mbox{.}(2020)]%
        {kong2020hifi}
\bibfield{author}{\bibinfo{person}{Jungil Kong}, \bibinfo{person}{Jaehyeon Kim}, {and} \bibinfo{person}{Jaekyoung Bae}.} \bibinfo{year}{2020}\natexlab{}.
\newblock \showarticletitle{Hi{F}i-{GAN}: Generative adversarial networks for efficient and high fidelity speech synthesis}.
\newblock \bibinfo{journal}{\emph{NeurIPS}} (\bibinfo{year}{2020}).
\newblock


\bibitem[Kumar et~al\mbox{.}(2019)]%
        {Kumar2019MelGANGA}
\bibfield{author}{\bibinfo{person}{Kundan Kumar}, \bibinfo{person}{Rithesh Kumar}, \bibinfo{person}{Thibault de Boissi{\`e}re}, {et~al\mbox{.}}} \bibinfo{year}{2019}\natexlab{}.
\newblock \showarticletitle{MelGAN: Generative Adversarial Networks for Conditional Waveform Synthesis}. In \bibinfo{booktitle}{\emph{NeurIPS}}.
\newblock


\bibitem[Lee et~al\mbox{.}(2022)]%
        {lee2022direct}
\bibfield{author}{\bibinfo{person}{Ann Lee}, \bibinfo{person}{Peng-Jen Chen}, \bibinfo{person}{Changhan Wang}, {et~al\mbox{.}}} \bibinfo{year}{2022}\natexlab{}.
\newblock \showarticletitle{Direct Speech-to-Speech Translation With Discrete Units}. In \bibinfo{booktitle}{\emph{ACL}}. \bibinfo{pages}{3327--3339}.
\newblock


\bibitem[Li et~al\mbox{.}(2023)]%
        {Li2023MERTAM}
\bibfield{author}{\bibinfo{person}{Yizhi Li}, \bibinfo{person}{Ruibin Yuan}, \bibinfo{person}{Ge Zhang}, {et~al\mbox{.}}} \bibinfo{year}{2023}\natexlab{}.
\newblock \showarticletitle{MERT: Acoustic Music Understanding Model with Large-Scale Self-supervised Training}.
\newblock \bibinfo{journal}{\emph{ArXiv}}  \bibinfo{volume}{abs/2306.00107} (\bibinfo{year}{2023}).
\newblock


\bibitem[Liu et~al\mbox{.}(2021)]%
        {Liu2021DiffSingerSV}
\bibfield{author}{\bibinfo{person}{Jinglin Liu}, \bibinfo{person}{Chengxi Li}, \bibinfo{person}{Yi Ren}, {et~al\mbox{.}}} \bibinfo{year}{2021}\natexlab{}.
\newblock \showarticletitle{DiffSinger: Singing Voice Synthesis via Shallow Diffusion Mechanism}. In \bibinfo{booktitle}{\emph{AAAI}}.
\newblock


\bibitem[Lu et~al\mbox{.}(2020)]%
        {Lu2020XiaoiceSingAH}
\bibfield{author}{\bibinfo{person}{Peiling Lu}, \bibinfo{person}{Jie Wu}, \bibinfo{person}{Jian Luan}, {et~al\mbox{.}}} \bibinfo{year}{2020}\natexlab{}.
\newblock \showarticletitle{XiaoiceSing: A High-Quality and Integrated Singing Voice Synthesis System}. In \bibinfo{booktitle}{\emph{Interspeech}}.
\newblock


\bibitem[Mustafa et~al\mbox{.}(2020)]%
        {Mustafa2020StyleMelGANAE}
\bibfield{author}{\bibinfo{person}{Ahmed Mustafa}, \bibinfo{person}{Nicola Pia}, {and} \bibinfo{person}{Guillaume Fuchs}.} \bibinfo{year}{2020}\natexlab{}.
\newblock \showarticletitle{Style{MelGAN}: An Efficient High-Fidelity Adversarial Vocoder with Temporal Adaptive Normalization}.
\newblock \bibinfo{journal}{\emph{ICASSP}} (\bibinfo{year}{2020}), \bibinfo{pages}{6034--6038}.
\newblock


\bibitem[Oura et~al\mbox{.}(2010)]%
        {sinsy}
\bibfield{author}{\bibinfo{person}{Keiichiro Oura}, \bibinfo{person}{Ayami Mase}, \bibinfo{person}{Tomohiko Yamada}, {et~al\mbox{.}}} \bibinfo{year}{2010}\natexlab{}.
\newblock \showarticletitle{{Recent development of the HMM-based singing voice synthesis system—Sinsy}}. In \bibinfo{booktitle}{\emph{Seventh ISCA Workshop on Speech Synthesis}}.
\newblock


\bibitem[Polyak et~al\mbox{.}(2021)]%
        {polyak21_interspeech}
\bibfield{author}{\bibinfo{person}{Adam Polyak}, \bibinfo{person}{Yossi Adi}, \bibinfo{person}{Jade Copet}, \bibinfo{person}{Eugene Kharitonov}, \bibinfo{person}{Kushal Lakhotia}, \bibinfo{person}{Wei-Ning Hsu}, \bibinfo{person}{Abdelrahman Mohamed}, {and} \bibinfo{person}{Emmanuel Dupoux}.} \bibinfo{year}{2021}\natexlab{}.
\newblock \showarticletitle{{Speech Resynthesis from Discrete Disentangled Self-Supervised Representations}}. In \bibinfo{booktitle}{\emph{Interspeech}}.
\newblock


\bibitem[Shi et~al\mbox{.}(2020)]%
        {Shi2020SequenceToSequenceSV}
\bibfield{author}{\bibinfo{person}{Jiatong Shi}, \bibinfo{person}{Shuai Guo}, \bibinfo{person}{Nan Huo}, {et~al\mbox{.}}} \bibinfo{year}{2020}\natexlab{}.
\newblock \showarticletitle{Sequence-To-Sequence Singing Voice Synthesis With Perceptual Entropy Loss}.
\newblock \bibinfo{journal}{\emph{ICASSP}} (\bibinfo{year}{2020}).
\newblock


\bibitem[Shi et~al\mbox{.}(2022)]%
        {shi2022muskits}
\bibfield{author}{\bibinfo{person}{Jiatong Shi}, \bibinfo{person}{Shuai Guo}, \bibinfo{person}{Tao Qian}, {et~al\mbox{.}}} \bibinfo{year}{2022}\natexlab{}.
\newblock \showarticletitle{Muskits: an End-to-End Music Processing Toolkit for Singing Voice Synthesis}. In \bibinfo{booktitle}{\emph{Interspeech}}.
\newblock


\bibitem[Shi et~al\mbox{.}(2023a)]%
        {shi2023bridging}
\bibfield{author}{\bibinfo{person}{Jiatong Shi}, \bibinfo{person}{Chan-Jan Hsu}, \bibinfo{person}{Holam Chung}, \bibinfo{person}{Dongji Gao}, \bibinfo{person}{Paola Garcia}, \bibinfo{person}{Shinji Watanabe}, \bibinfo{person}{Ann Lee}, {and} \bibinfo{person}{Hung-yi Lee}.} \bibinfo{year}{2023}\natexlab{a}.
\newblock \showarticletitle{Bridging speech and textual pre-trained models with unsupervised ASR}. In \bibinfo{booktitle}{\emph{ICASSP}}.
\newblock


\bibitem[Shi et~al\mbox{.}(2023b)]%
        {shi2023multi}
\bibfield{author}{\bibinfo{person}{Jiatong Shi}, \bibinfo{person}{Hirofumi Inaguma}, \bibinfo{person}{Xutai Ma}, {et~al\mbox{.}}} \bibinfo{year}{2023}\natexlab{b}.
\newblock \showarticletitle{Multi-resolution {HuBERT}: Multi-resolution Speech Self-Supervised Learning with Masked Unit Prediction}. In \bibinfo{booktitle}{\emph{ICLR}}.
\newblock


\bibitem[Shi et~al\mbox{.}(2023c)]%
        {shi2023enhancing}
\bibfield{author}{\bibinfo{person}{Jiatong Shi}, \bibinfo{person}{Yun Tang}, \bibinfo{person}{Ann Lee}, \bibinfo{person}{Hirofumi Inaguma}, \bibinfo{person}{Changhan Wang}, \bibinfo{person}{Juan Pino}, {and} \bibinfo{person}{Shinji Watanabe}.} \bibinfo{year}{2023}\natexlab{c}.
\newblock \showarticletitle{Enhancing Speech-To-Speech Translation with Multiple TTS Targets}. In \bibinfo{booktitle}{\emph{ICASSP}}.
\newblock


\bibitem[Tang et~al\mbox{.}(2024)]%
        {tang2024singomd}
\bibfield{author}{\bibinfo{person}{Yuxun Tang}, \bibinfo{person}{Yuning Wu}, \bibinfo{person}{Jiatong Shi}, {and} \bibinfo{person}{Qin Jin}.} \bibinfo{year}{2024}\natexlab{}.
\newblock \showarticletitle{SingOMD: Singing Oriented Multi-resolution Discrete Representation Construction from Speech Models}. In \bibinfo{booktitle}{\emph{Interspeech}}.
\newblock


\bibitem[Wang et~al\mbox{.}(2023)]%
        {wang2023neural}
\bibfield{author}{\bibinfo{person}{Chengyi Wang}, \bibinfo{person}{Sanyuan Chen}, \bibinfo{person}{Yu Wu}, \bibinfo{person}{Ziqiang Zhang}, \bibinfo{person}{Long Zhou}, \bibinfo{person}{Shujie Liu}, \bibinfo{person}{Zhuo Chen}, \bibinfo{person}{Yanqing Liu}, \bibinfo{person}{Huaming Wang}, \bibinfo{person}{Jinyu Li}, {et~al\mbox{.}}} \bibinfo{year}{2023}\natexlab{}.
\newblock \showarticletitle{Neural codec language models are zero-shot text to speech synthesizers}.
\newblock \bibinfo{journal}{\emph{arXiv preprint arXiv:2301.02111}} (\bibinfo{year}{2023}).
\newblock


\bibitem[Wang et~al\mbox{.}(2022b)]%
        {wang2022xiaoicesing}
\bibfield{author}{\bibinfo{person}{Chunhui Wang}, \bibinfo{person}{Chang Zeng}, {and} \bibinfo{person}{Xing He}.} \bibinfo{year}{2022}\natexlab{b}.
\newblock \showarticletitle{Xiaoicesing 2: A high-fidelity singing voice synthesizer based on generative adversarial network}.
\newblock \bibinfo{journal}{\emph{arXiv preprint arXiv:2210.14666}} (\bibinfo{year}{2022}).
\newblock


\bibitem[Wang et~al\mbox{.}(2022a)]%
        {Wang2022SingingTacotronGD}
\bibfield{author}{\bibinfo{person}{Tao Wang}, \bibinfo{person}{Ruibo Fu}, \bibinfo{person}{Jiangyan Yi}, {et~al\mbox{.}}} \bibinfo{year}{2022}\natexlab{a}.
\newblock \showarticletitle{Singing-{T}acotron: Global Duration Control Attention and Dynamic Filter for End-to-end Singing Voice Synthesis}.
\newblock \bibinfo{journal}{\emph{Proceedings of the 1st International Workshop on Deepfake Detection for Audio Multimedia}} (\bibinfo{year}{2022}).
\newblock


\bibitem[Watanabe et~al\mbox{.}(2018)]%
        {Watanabe2018ESPnetES}
\bibfield{author}{\bibinfo{person}{Shinji Watanabe}, \bibinfo{person}{Takaaki Hori}, \bibinfo{person}{Shigeki Karita}, \bibinfo{person}{Tomoki Hayashi}, \bibinfo{person}{Jiro Nishitoba}, \bibinfo{person}{Yuya Unno}, \bibinfo{person}{Nelson Yalta}, \bibinfo{person}{Jahn Heymann}, \bibinfo{person}{Matthew Wiesner}, \bibinfo{person}{Nanxin Chen}, \bibinfo{person}{Adithya Renduchintala}, {and} \bibinfo{person}{Tsubasa Ochiai}.} \bibinfo{year}{2018}\natexlab{}.
\newblock \showarticletitle{ESPnet: End-to-End Speech Processing Toolkit}. In \bibinfo{booktitle}{\emph{Interspeech}}.
\newblock


\bibitem[Wu et~al\mbox{.}(2023a)]%
        {wu2023phoneix}
\bibfield{author}{\bibinfo{person}{Yuning Wu}, \bibinfo{person}{Jiatong Shi}, \bibinfo{person}{Tao Qian}, \bibinfo{person}{Dongji Gao}, {and} \bibinfo{person}{Qin Jin}.} \bibinfo{year}{2023}\natexlab{a}.
\newblock \showarticletitle{{PHONEix}: Acoustic Feature Processing Strategy for Enhanced Singing Pronunciation With Phoneme Distribution Predictor}.
\newblock \bibinfo{journal}{\emph{ICASSP}}.
\newblock


\bibitem[Wu et~al\mbox{.}(2023b)]%
        {Wu2023ASE}
\bibfield{author}{\bibinfo{person}{Yuning Wu}, \bibinfo{person}{Yifeng Yu}, \bibinfo{person}{Jiatong Shi}, \bibinfo{person}{Tao Qian}, {and} \bibinfo{person}{Qin Jin}.} \bibinfo{year}{2023}\natexlab{b}.
\newblock \showarticletitle{A Systematic Exploration of Joint-training for Singing Voice Synthesis}.
\newblock \bibinfo{journal}{\emph{ArXiv}}  \bibinfo{volume}{abs/2308.02867} (\bibinfo{year}{2023}).
\newblock


\bibitem[Wu et~al\mbox{.}(2024)]%
        {wu2024toksing}
\bibfield{author}{\bibinfo{person}{Yuning Wu}, \bibinfo{person}{Chunlei zhang}, \bibinfo{person}{Jiatong Shi}, \bibinfo{person}{Yuxun Tang}, \bibinfo{person}{Shan Yang}, {and} \bibinfo{person}{Qin Jin}.} \bibinfo{year}{2024}\natexlab{}.
\newblock \showarticletitle{TokSing: Singing Voice Synthesis based on Discrete Tokens}. In \bibinfo{booktitle}{\emph{Interspeech}}.
\newblock


\bibitem[Yamamoto et~al\mbox{.}(2023)]%
        {nnsvs}
\bibfield{author}{\bibinfo{person}{Ryuichi Yamamoto}, \bibinfo{person}{Reo Yoneyama}, {and} \bibinfo{person}{Tomoki Toda}.} \bibinfo{year}{2023}\natexlab{}.
\newblock \showarticletitle{NNSVS: A neural network-based singing voice synthesis toolkit}. In \bibinfo{booktitle}{\emph{ICASSP}}.
\newblock


\bibitem[Yan et~al\mbox{.}(2023)]%
        {yan-etal-2023-espnet}
\bibfield{author}{\bibinfo{person}{Brian Yan}, \bibinfo{person}{Jiatong Shi}, \bibinfo{person}{Yun Tang}, \bibinfo{person}{Hirofumi Inaguma}, \bibinfo{person}{Yifan Peng}, \bibinfo{person}{Siddharth Dalmia}, \bibinfo{person}{Peter Pol{\'a}k}, \bibinfo{person}{Patrick Fernandes}, \bibinfo{person}{Dan Berrebbi}, \bibinfo{person}{Tomoki Hayashi}, \bibinfo{person}{Xiaohui Zhang}, \bibinfo{person}{Zhaoheng Ni}, \bibinfo{person}{Moto Hira}, \bibinfo{person}{Soumi Maiti}, \bibinfo{person}{Juan Pino}, {and} \bibinfo{person}{Shinji Watanabe}.} \bibinfo{year}{2023}\natexlab{}.
\newblock \showarticletitle{{ESP}net-{ST}-v2: Multipurpose Spoken Language Translation Toolkit}. In \bibinfo{booktitle}{\emph{ACL}}.
\newblock


\bibitem[Yang et~al\mbox{.}(2024)]%
        {yang2023towards}
\bibfield{author}{\bibinfo{person}{Yifan Yang}, \bibinfo{person}{Feiyu Shen}, \bibinfo{person}{Chenpeng Du}, \bibinfo{person}{Ziyang Ma}, \bibinfo{person}{Kai Yu}, \bibinfo{person}{Daniel Povey}, {and} \bibinfo{person}{Xie Chen}.} \bibinfo{year}{2024}\natexlab{}.
\newblock \showarticletitle{Towards Universal Speech Discrete Tokens: A Case Study for {ASR} and {TTS}}. In \bibinfo{booktitle}{\emph{Proc. ICASSP}}.
\newblock


\bibitem[Yu et~al\mbox{.}(2024)]%
        {yu2024visinger2}
\bibfield{author}{\bibinfo{person}{Yifeng Yu}, \bibinfo{person}{Jiatong Shi}, \bibinfo{person}{Yuning Wu}, {and} \bibinfo{person}{Shinji Watanabe}.} \bibinfo{year}{2024}\natexlab{}.
\newblock \showarticletitle{VISinger2+: End-to-End Singing Voice Synthesis Augmented by Self-Supervised Learning Representation}.
\newblock \bibinfo{journal}{\emph{ArXiv}}  \bibinfo{volume}{abs/2406.08761} (\bibinfo{year}{2024}).
\newblock


\bibitem[Yuxun~Tang(2024)]%
        {singmos}
\bibfield{author}{\bibinfo{person}{Yuning Wu Qin~Jin Yuxun~Tang, Jiatong~Shi}.} \bibinfo{year}{2024}\natexlab{}.
\newblock \showarticletitle{SingMOS: An extensive Open-Source Singing Voice Dataset for MOS Prediction}.
\newblock \bibinfo{journal}{\emph{arXiv preprint arXiv:2406.10911}} (\bibinfo{year}{2024}).
\newblock


\bibitem[Zeghidour et~al\mbox{.}(2021)]%
        {Zeghidour2021SoundStreamAE}
\bibfield{author}{\bibinfo{person}{Neil Zeghidour}, \bibinfo{person}{Alejandro Luebs}, \bibinfo{person}{Ahmed Omran}, {et~al\mbox{.}}} \bibinfo{year}{2021}\natexlab{}.
\newblock \showarticletitle{SoundStream: An End-to-End Neural Audio Codec}.
\newblock \bibinfo{journal}{\emph{TASLP}}  \bibinfo{volume}{30} (\bibinfo{year}{2021}), \bibinfo{pages}{495--507}.
\newblock


\bibitem[Zhang et~al\mbox{.}(2023)]%
        {zhang2023speechgpt}
\bibfield{author}{\bibinfo{person}{Dong Zhang}, \bibinfo{person}{Shimin Li}, \bibinfo{person}{Xin Zhang}, \bibinfo{person}{Jun Zhan}, \bibinfo{person}{Pengyu Wang}, \bibinfo{person}{Yaqian Zhou}, {and} \bibinfo{person}{Xipeng Qiu}.} \bibinfo{year}{2023}\natexlab{}.
\newblock \showarticletitle{SpeechGPT: Empowering Large Language Models with Intrinsic Cross-Modal Conversational Abilities}. In \bibinfo{booktitle}{\emph{The 2023 Conference on Empirical Methods in Natural Language Processing}}.
\newblock


\bibitem[Zhang et~al\mbox{.}(2021)]%
        {Zhang2021VISingerVI}
\bibfield{author}{\bibinfo{person}{Yongmao Zhang}, \bibinfo{person}{Jian Cong}, \bibinfo{person}{Heyang Xue}, {et~al\mbox{.}}} \bibinfo{year}{2021}\natexlab{}.
\newblock \showarticletitle{{VIS}inger: Variational Inference with Adversarial Learning for End-to-End Singing Voice Synthesis}.
\newblock \bibinfo{journal}{\emph{ICASSP}} (\bibinfo{year}{2021}), \bibinfo{pages}{7237--7241}.
\newblock


\bibitem[Zhang et~al\mbox{.}(2022)]%
        {Zhang2022VISinger2H}
\bibfield{author}{\bibinfo{person}{Yongmao Zhang}, \bibinfo{person}{Heyang Xue}, \bibinfo{person}{Hanzhao Li}, {et~al\mbox{.}}} \bibinfo{year}{2022}\natexlab{}.
\newblock \showarticletitle{{VIS}inger 2: High-Fidelity End-to-End Singing Voice Synthesis Enhanced by Digital Signal Processing Synthesizer}.
\newblock \bibinfo{journal}{\emph{ArXiv}}  \bibinfo{volume}{abs/2211.02903} (\bibinfo{year}{2022}).
\newblock


\end{thebibliography}

\end{document}